\documentstyle[osa,manuscript]{revtex}
\title{A new method for detection of exciton Bose condensation using stimulated two-photon emission}
\author{Yu.~E.~Lozovik and A~V.~Poushnov\footnote{E-mail: poushnov@isan.troitsk.ru}}
\address{Institute of Spectroscopy, Russian Academy of Sciences, 142092 Troitsk, Moscow Region, Russia}

\begin{document}
\maketitle

\centerline{Submitted {18 August, 1998}}

\centerline{Zh. \'Eksp. Teor. Fiz. {\bf 115}, (April 1999)}

\begin{abstract}
Stimulated two-photon emission by Bose-condensed excitons accompanied by a
coherent two-exciton recombination, i.e., by simultaneous recombination of two
excitons with opposite momenta leaving unchanged the occupation numbers of
excitonic states with momenta ${\bf p}\ne 0$, is investigated.
Raman light scattering (RLS) accompanied by a similar two-exciton
recombination (or generation of two excitons) is also analyzed. The processes
under consideration can occur only if a system contains Bose condensate,
therefore, their detection can be used as a new method to reveal Bose
condensation of excitons. The
recoil momentum, which corresponds to a change in the momentum of the
electromagnetic field in the processes, is transferred to phonons or
impurities. If the recoil momentum is transmitted to optical phonons with
frequency $\omega_0^s$, whose occupation numbers are negligible, and the
incident light frequency $\omega<2\Omega_-$, where
$\Omega_-=\Omega-\omega_0^s$ and $\Omega$ is the light frequency
corresponding to the
recombination of an exciton with zero momentum, the stimulated two-photon
emission and RLS with the coherent two-exciton recombination lead to the
appearance of a line at $2\Omega_--\omega$ and an anti-Stokes component at
$\omega+2\Omega_-$, respectively. At $\omega>2\Omega_-$ the RLS spectrum
contains Stokes
and anti-Stokes components at frequencies $\omega\pm2\Omega_-$, whereas the
stimulated two-photon emission is impossible. Formulas for the cross
sections at finite temperatures are obtained for the processes
under consideration. Our estimates
indicate that a spectral line at $2\Omega_--\omega$, corresponding to the
stimulated two-photon emission accompanied by the coherent optical
phonon-assisted two-exciton recombination can be experimentally detected in
Cu$_2$O.
\end{abstract}

\section{Introduction}

The most interesting collective effects in systems of excitons are the
anticipated exciton Bose condensation and superfluidity (see Refs.~1-7 and
references therein). Recently a number of publications reported on
the detection of Bose condensation and superfluidity of excitons
in Cu$_2$O based on observations of changes in exciton luminescence
spectra\cite{8,9}
and ballistic transport of excitons,\cite{9,10,11} which have been
discussed in literature.\cite{12,13,14} Observations of condensation of
indirect excitons in coupled quantum wells under strong magnetic fields
have also been reported (see Ref.~15, a theoretical discussion in
Refs.~16-18, and references therein). In this connection, the detailed
investigation of coherent exciton properties, whose
detection could be used to reveal exciton Bose condensation, seems to be
important.

If a system of excitons is in a Bose condensed state, the mean values of
the annihilation (creation) operator of the exciton with zero momentum in
the ground state are not zero:
\begin{eqnarray}
\label{1}
\langle N-1|Q_0|N\rangle=\langle N+1|Q^+_0|N\rangle=\sqrt{N_0}.
\end{eqnarray}
Here $|N\rangle$ is the ground state of the excitonic system with the
average number of excitons $N$, $Q_0$ is the annihilation operator of an
exciton with zero momentum, and $N_0$ is the number of
excitons in the condensate.

Equation~(1) clearly shows that, as a result of the recombination
(generation) of an exciton with zero momentum, a system of Bose-condensed
excitons transfers to the ground state, that differs from the initial one in
the average number of excitons with momentum ${\bf p}=0$. The
recombination of excitons with zero momentum leads to the appearance of a
peak (the so-called condensate peak) in the excitonic luminescence spectrum
at frequency $\Omega=[E_0(N)-E_0(N-1)]/\hbar$, where $E_0(N)$ is the energy
of the ground state of the excitonic system.

If the exciton-exciton interaction is nonvanishing, then, in addition to the
mean values defined by Eq.~(1), products of two annihilation (creation)
operators of excitons with opposite momenta averaged over the ground
state of the Bose-condensed excitonic system (the so-called anomalous
averages) are not zero:
\begin{eqnarray}
\label{2}
\langle N-2|Q_{-p}Q_{p}|N\rangle\neq 0,\quad
\langle N+2|Q^+_{-p}Q^+_{p}|N\rangle\neq 0.
\end{eqnarray}

The unusual optical properties inherent in Bose-condensed state of
interacting excitons due to nonvanishing anomalous means~(2) are considered
in this paper. It will be shown that due to the interaction with the
electromagnetic field, the coherent recombination (or generation), i.e.,
the simultaneous recombination (or creation) of two excitons with opposite
momenta, corresponding to anomalous averages~(2) is possible. In such
processes, the occupation numbers of excitons with ${\bf p}\ne 0$ are
unchanged, and the final state of excitons differs from the initial one
only in the average number of excitons with zero momentum. In particular,
after the two-exciton recombination, the average number of condensate
excitons is reduced by two.

The coherent two-exciton recombination can contribute, for example, to the
stimulated two-photon emission or to Raman light scattering (RLS) by
Bose-condensed excitons. RLS can also be accompanied by the coherent
generation of two excitons. In these processes, the momentum of the
exciton-photon system is not conserved: the recoil momentum, equal to the
change in the momentum of the electromagnetic field, is transferred to
phonons or impurities.\cite{19,20} In this paper we consider the processes in
which
the recoil momentum is transferred to two optical phonons. Such processes
seem to be most probable in the excitonic system in Cu$_2$O crystal, which is
one of most interesting crystals in view of the observation of exciton Bose
condensation. In fact, the radiative recombination accompanied by the
transmission of the recoil momentum to one optical phonon is typical for
excitons in Cu$_2$O.\cite{7} Using the energy and momentum conservation laws,
one can prove that, in a defect-free crystal, the coherent recombination of
two excitons is possible only if the recoil momentum is transferred to two
phonons.

At low temperatures, the occupation numbers of optical phonons are small,
therefore, it is most probable that the recoil momentum is transferred to two
phonons
generated in the process. If the phonon dispersion is negligible and the
incident light frequency $\omega<2\Omega_-$, a line in the spectrum of the
stimulated two-phonon emission at $2\Omega_--\omega$ and an anti-Stokes
component in the RLS spectrum at $\omega+2\Omega_-$ should appear. Here
$\Omega_-=\Omega-\omega_0^s$ and $\omega_0^s$ is the optical phonon
frequency. Both these lines correspond to the coherent two-exciton
recombination: the energy of the initial state of the system is higher than
the energy of its final state by $2\hbar\Omega$, where $\Omega$ is the
frequency corresponding to the recombination of an exciton with zero
momentum. If $\omega>2\Omega_-$, the RLS spectrum should contain the
anti-Stokes component at $\omega+2\Omega_-$, which corresponds to the coherent
two-exciton recombination, and the Stokes component at $\omega-2\Omega_-$
due to the coherent generation of two excitons. The stimulated emission of
two photons is impossible in this case. The appearance of the lines at
frequencies $|\omega\pm 2\Omega_-|$ is possible only if the excitons are in
the Bose-condensed state, and after a transition to the normal state these
lines should disappear.

The paper is organized as follows. In Section~2 the stimulated two-photon
emission with the coherent two-exciton recombination accompanied by the
transmission of the recoil momentum to phonons is considered. The diagram
technique is used to obtain the cross sections of two-photon processes
involving the coherent two-exciton recombination (or generation) at finite
temperatures. This approach allows one to express the appropriate elements
of the $S$-matrix in a natural manner in terms of anomalous Green's
functions of Bose-condensed excitons. The cross section of the stimulated
two-photon emission with the coherent phonon-assisted two-exciton
recombination is obtained, and its temperature dependence is studied. It
turns out that this dependence can be nonmonotonic under certain conditions.
Namely, in a certain temperature interval below $T_{\rm c}$ the cross section
of the stimulated two-photon emission can increase with the growth of
temperature  and can become even higher than it is at $T=0$. The causes of
this unusual temperature dependence is investigated.

Section~3 is dedicated to RLS accompanied by the coherent processes of
two-exciton recombination or generation. In Section~4 the possibility
of the experimental observation of the lines at frequencies
$|\omega\pm 2\Omega_-|$ corresponding to the stimulated two-photon emission
and RLS is analyzed. Our numerical estimates for excitons in Cu$_2$O
indicate that a spectral line at $2\Omega_--\omega$ corresponding to the
stimulated optical phonon-assisted two-exciton recombination can be detected
and, therefore, can be used to reveal exciton Bose condensation.

\section{Stimulated two-photon emission accompanied by coherent
two-exciton recombination}

The effective Hamiltonian describing phonon-assisted radiative
recombination (generation) of excitons can be expressed as follows (see
Ref.~20 and Appendix A):
\begin{eqnarray}
\label{3}
\hat H_L&=&\sum_{pq}^{}\left[
L^>_{pq}e^{-i\Omega t}Q_p(t)c^+_{q}(t)b^+_{p-q}(t)+
L^<_{pq}e^{-i\Omega t}Q_p(t)c^+_{q}(t)b_{q-p}(t)
\right.\nonumber\\
&+& \left.
L'^>_{pq}e^{-i\Omega t}Q_p(t)c_{q}(t)b^+_{p+q}(t)+
L'^<_{pq}e^{-i\Omega t}Q_p(t)c_{q}(t)b_{-p-q}(t)+{\rm H.c.}\right],
\end{eqnarray}
where
\begin{displaymath}
L^{>(<)}_{pq}=i\sqrt{2\pi\omega_q}{\bf e}^*{\bf f}^{>(<)}_{pq},\quad
L'^{>(<)}_{pq}=-i\sqrt{2\pi\omega_q}{\bf ef}'^{>(<)}_{pq},
\end{displaymath}
$\Omega$ is the frequency corresponding to the recombination of an exciton
with zero momentum. Hamiltonian~(3) is written in the Heisenberg
representation. Here $Q_p(t)=Q_p\exp[-i\epsilon(p)t]$ and
$b_p(t)=b_p\exp(-i\omega_p^st)$ are the annihilation operators of an
exciton and a phonon with momentum $p$, respectively,
$c_q(t)=c_q\exp(-i\omega_qt)$ is the annihilation operator of a photon with
momentum $q$ ($\omega_q$ and ${\bf e}$ are the photon frequency and its
polarization unit vector). The exciton energy is measured with respect to
the bottom of the exciton band: $\epsilon(0)=0$. The effective matrix
elements ${\bf f}_{pq}^{>(<)}$ and ${\bf f}'^{>(<)}_{pq}$ are responsible for
the recombination of an exciton with momentum ${\bf p}$, which includes, in
addition to the emission (absorption) of a photon with momentum ${\bf q}$,
the simultaneous emission or absorption of a phonon\footnote{In a general
case, the radiative recombination of an exciton can result in emission
(absorption) of an arbitrary number of phonons. When using Hamiltonian~(3),
we limit our analysis for simplicity to the case of excitonic recombination
with emission (absorption) of one phonon.} (see Ref.~20 and Appendix~A).

By expanding the evolution operator
\begin{displaymath}
\hat S(t)=T_t{\rm exp}
\left[-i\int\limits_{-\infty}^{t}\hat H_L(t')dt'
\right]
\end{displaymath}
in powers of $\hat{H}_L$ and retaining terms of up to the second order, we
obtain an expression for the elements of the $S$-matrix corresponding to
phonon-assisted two-photon processes:
\begin{equation}
\label{4}
S_{n'n}=\frac{(-i)^2}{2!}
\mathop{\int\!\!\int}\limits_{-\infty\quad }^{\quad\infty}
\langle n'|T_t\hat H_L(t')\hat H_L(t'')|n\rangle dt'dt'',
\end{equation}
where $n$ and $n'$ label the initial and final states of the system
composed of excitons and phonons + electromagnetic field.

Let us consider the two-photon emission by excitons in the Bose condensed
state
due to the coherent two-exciton recombination, i.e., a transition of the
excitonic system from state $|n\rangle_{\rm exc}=|n,N\rangle_{\rm exc}$ to
the state $|m\rangle_{\rm exc}=|n,N-2\rangle_{\rm exc}$, which differs
from the initial state in the average number of excitons with momentum
${\bf p}=0$. The change in the electromagnetic field momentum is ${\bf k'+k}$
in this process, where ${\bf k}$ and ${\bf k'}$ are the momenta of emitted
photons. The recoil momentum $\delta{\bf k}=-({\bf k'+k})$ is entirely
transferred to phonons since the momentum of the excitonic system is zero in
both the initial and final states.

For the element of the $S$-matrix corresponding to the coherent
phonon-assisted two-exciton recombination, we have
\begin{eqnarray}
\label{5}
(S_p)_{mn}&=&-\frac{1}{2}
\mathop{\int\!\!\int}\limits_{-\infty\quad}^{\quad\infty}
dt'dt''\exp\left[{-i\Omega(t'+t'')}\right]
\nonumber \\
&\times& \left\{\left[
L^>_{pk}L^>_{-pk'}\langle m|T_tQ_p(t')Q_{-p}(t'')|n\rangle_{\rm exc}
\langle f|T_tb^+_{p-k}(t')b^+_{-p-k'}(t'')|i\rangle_{\rm phon}
\right.\right.\nonumber\\
&+& \left.\left.
L^>_{q-p,k}L^>_{p-q,k'}\langle m|T_tQ_{q-p}(t')Q_{p-q}(t'')|n\rangle_{\rm exc}
\langle f|T_tb^+_{-p-k'}(t')b^+_{p-k}(t'')|i\rangle_{\rm phon}\right]
\right.\nonumber\\
&\times&\left.
\langle f|T_tc^+_{k}(t')c^+_{k'}(t'')|i\rangle_{\rm phot}
\right. \nonumber \\
&+& \left. \left[L^>_{-pk'}L^>_{pk}\langle m|T_tQ_{-p}(t')Q_{p}(t'')
|n\rangle_{\rm exc}\langle f|T_tb^+_{-p-k'}(t')b^+_{p-k}(t'')|i\rangle_{\rm phon}
\right. \right. \nonumber \\
&+& \left.\left.L^>_{p-q,k'}L^>_{q-p,k}
\langle m|T_tQ_{p-q}(t')Q_{q-p}(t'')|n\rangle_{\rm exc}
\langle f|T_tb^+_{p-k}(t')b^+_{-p-k'}(t'')|i\rangle_{\rm phon}\right]
\right.\nonumber\\
&\times& \left.\langle f|T_tc^+_{k'}(t')c^+_{k}(t'')|i\rangle_{\rm phot}\right\},
\end{eqnarray}
where ${\bf q=k-k'}$. Here $|i\rangle_{\rm phot}=|0\rangle_{\rm phot}$ and
$|f\rangle_{\rm phot}=|1_k,1_{k'}\rangle_{\rm phot}$ are the initial and
final states of the electromagnetic field, respectively. Assuming that the
phonons are optical and the lattice temperature $T_{\rm lat}$, which is,
generally speaking, different from the excitonic temperature $T$, is
sufficiently small ($T_{\rm lat}\ll\omega_0^s$, where $\omega_0^s$ is the
characteristic energy of optical phonons), we suppose that $|i\rangle_{\rm
phon}=|0\rangle_{\rm phon}$ and $|f\rangle=|1_{p-k},1_{-p-k'}\rangle_{\rm
phon}$.

By performing averaging over the Gibbs distribution for the excitonic
system, we obtain the element of the $S$-matrix responsible for the
two-photon emission that transforms the system from its state of
thermodynamic equilibrium $|i\rangle_{\rm
exc}=\sum\limits_n\exp[(F-E_n(N)+\mu N)/T]|n,N\rangle_{\rm exc}$ to
$|f\rangle_{\rm exc}=Q^2_0|i\rangle/N_0$:
\begin{eqnarray}
\label{6}
(S_p)_{fi}=\sum_{n}^{}{\rm exp}[(F-E_n(N)+\mu N)/T](S_p)_{mn}.
\end{eqnarray}
By expressing the $S$-matrix element~(6) in terms of the anomalous Green's
function of excitons, we obtain
\begin{eqnarray}
\label{7}
(S_p)_{fi}&=&-\frac{1}{2}
\mathop{\int\!\!\int}\limits_{-\infty\quad}^{\quad\infty}
dt'dt''\exp\left[{-i\Omega(t'+t'')}\right]
\nonumber \\
&\times& \left\{\left[L^>_{pk}L^>_{-pk'}
\left(n_0(T)\delta_p+i\hat G_{-p}(t'-t'')\right)
\langle f|T_tb^+_{p-k}(t')b^+_{-p-k'}(t'')|i\rangle_{\rm phon}
\right.\right.
\nonumber\\
&+& \left. \left.
L^>_{q-p,k}L^>_{p-q,k'}\left(n_0(T)\delta_{p-q}+i\hat G_{p-q}(t'-t'')\right)
\langle f|T_tb^+_{-p-k'}(t')b^+_{p-k}(t'')|i\rangle_{\rm phon}\right]
\right.
\nonumber\\
&\times& \left.
\langle f|T_tc^+_{k}(t')c^+_{k'}(t'')|i\rangle_{\rm phot}
\right.
\nonumber \\
&+& \left. \left[
L^>_{-pk'}L^>_{pk}\left(n_0(T)\delta_p+i\hat G_p(t'-t'')\right)
\langle f|T_tb^+_{-p-k'}(t')b^+_{p-k}(t'')|i\rangle_{\rm phon}
\right. \right.
\nonumber \\
&+& \left. \left. L^>_{p-q,k'}L^>_{q-p,k}
\left(n_0(T)\delta_{p-q}+i\hat G_{q-p}(t'-t'')\right)
\langle f|T_tb^+_{p-k}(t')b^+_{-p-k'}(t'')|i\rangle_{\rm phon}\right]
\right.
\nonumber \\
&\times& \left.
\langle f|T_tc^+_{k'}(t')c^+_{k}(t'')|i\rangle_{\rm phot}\right\},
\end{eqnarray}
where $\delta_p=1$ at $p=0$ and $\delta_p=0$ at $p\ne 0$. Here
$\hat{G}_p(t'-t'')$ is the causal Green's function of Bose-condensed
excitons at temperature $T$:
\begin{eqnarray}
\label{8}
\hat G_p(t'-t'')&=&-i(1{-}\delta_p)
\nonumber\\
&\times&
\sum_{n}{\exp}\left[(F{-}E_n(N){+}\mu N)/T\right]
\langle n,N{-}2|T_tQ_{-p}(t')Q_p(t'')|n,N\rangle_{\rm exc},
\end{eqnarray}
and function $n_0(T)$ is the density of excitons in the condensate at this
temperature.

The resulting element~(7) of the $S$-matrix is expressed by the sum of
diagrams shown in Fig.~1. The lines with oppositely directed arrows denote
the causal anomalous Green's function of excitons in the Bose-condensed
state at $T>0$ (if the momenta next to this line vanish, it corresponds to
function $n_0(T)$). The wavy lines correspond to photon creation operators,
and the dashed lines indicate phonon creation operators. The vertices on
these diagrams
correspond to matrix elements $L_{pk}^>$, where $p$ and $k$ are the momenta
of the exciton and photon lines originating at the
vertex.\footnote{Calculations concerning the two-photon emission and RLS
under discussion could be performed using Keldysh's elegant diagrammatic
technique (see, e.g, Ref.~21, which is devoted to a problem that requires a
similar technique). In our opinion, however, our approach used in this
specific case is more visual.}

Integration with respect to $t'-t''$ and $t''$ yields
\begin{equation}
\label{9}
(S_p)_{fi}=2\pi iT_{k'k}({\bf p})
\left[(\sqrt{2}-1)\delta({\bf p}-{\bf q}/2)+1\right]
\delta(\omega'+\omega+\omega^s_{p-k}+\omega^s_{-p-k'}-2\Omega),
\end{equation}
where
\begin{eqnarray}
\label{10}
T_{k'k}({\bf p})&=&
i\left\{L^>_{pk}L^>_{-pk'}\left[2\pi n_0(T)\delta_p\delta(\omega+\omega^s_{p-k}
-\Omega)+i\hat G_{p}(\omega+\omega^s_{p-k}-\Omega)\right]
\right. \nonumber \\
&{+}& \left.
L^>_{q{-}p,k}L^>_{p{-}q,k'}
\left[2\pi n_0(T)\delta_{p{-}q}\delta(\omega+\omega^s_{-p-k'}{-}\Omega){+}
i\hat G_{p-q}(\omega{+}\omega^s_{-p-k'}{-}\Omega)\right]\right\}
\end{eqnarray}
is the matrix element of the two-photon emission due to the coherent
phonon-assisted two-exciton recombination, which is similar to the
scattering amplitude in the collision problem.\cite{22} In deriving this
equation, we have taken into account the fact that the anomalous Green's
function is an even function of frequency and does not depend on the momentum
direction. The sum in the brackets in Eq.~(9) takes into account the fact
that the momenta of emitted phonons are equal at ${\bf p}={\bf q}/2$.

Let us limit our discussion to the stimulated phonon-assisted two-photon
emission with a negligible dispersion of phonons ($\omega_q^s=\omega_0^s$).
It follows from Eq.~(9) that the stimulated two-photon emission of this kind
leads to the appearance of a line at frequency $2\Omega_--\omega$, where
$\Omega_-=\Omega-\omega_0^s$ and $\omega$ is the incident light
frequency.\footnote{In a general case, the number of phonons involved
in the process can be arbitrary. Moreover, the recoil momentum (the whole
or a fraction of it) can be transferred to impurities. Thus, the stimulated
two-photon emission can result in the appearance of the spectral lines at
frequencies $2(\Omega-n\omega_0^s)$, where $n$ is an arbitrary integer.}

The differential cross section of the stimulated two-photon emission
corresponding to the coherent phonon-assisted two-exciton recombination is
given by
\begin{equation}
\label{11}
d\sigma^L=\frac{2\pi}{c}
\left[\frac{1}{2}\sum_{p\neq q/2}^{}|T_{k'k}({\bf p})|^2+
2|T_{k'k}({\bf q}/2)|^2 \right]\frac{(2\Omega_--\omega)^2}{(2\pi c)^3}do',
\end{equation}
where
\begin{eqnarray}
\label{12}
T_{k'k}({\bf p})&=&i\left\{L^>_{pk}L^>_{-pk'}\left[2\pi n_0(T)\delta_p
\delta(\omega-\Omega_-)+i\hat G_{p}(\omega-\Omega_-)\right]
\right. \nonumber \\
&+& \left.
L^>_{q-p,k}L^>_{p-q,k'}\left[2\pi n_0(T)\delta_{p-q}\delta(\omega-\Omega_-)+
i\hat G_{p-q}(\omega-\Omega_-)\right]\right\}.
\end{eqnarray}
The factor $1/2$ in front of the sum over ${\bf p}$ in Eq.~(11) is
introduced because the sum over all possible ${\bf p}$ includes the
emission of two phonons with momenta ${\bf p-k}$ and ${\bf -p-k'}$ two
times: $T_{k'k}({\bf p})=T_{k'k}({\bf -p+q})$.

It is clear that at $\omega\ne\Omega_-$ the summands proportional to
$n_0(T)$ do not contribute to the cross section~(11). In this case, it is
proportional to anomalous Green's functions, which are determined, as is
well known, not only by the presence of Bose condensate, but also by the
interaction among particles. Thus, the stimulated two-photon emission
corresponding to the coherent phonon-assisted two-exciton recombination at
$\omega\ne\Omega_-$ can take place only in a non-ideal gas of excitons with
Bose condensate.

Assuming that the condition $\omega\ne\Omega_-$ is fulfilled, let us express
cross section~(11) as follows:
\begin{equation}
\label{13}
d\sigma^L=\frac{\omega(2\Omega_--\omega)^3}{c^4}
\left[\frac{1}{2}\sum_{p\neq q/2}|(s_p)_{nm}e'^*_ne^*_m|^2+
2|(s_{q/2})_{nm}e'^*_ne^*_m|^2\right]do',
\end{equation}
where
\begin{equation}
\label{14}
(s_p)_{nm}=\hat G_p(\omega-\Omega_-)(f^>_{-pk'})_n(f^>_{pk})_m+
\hat G_{p-q}(\omega-\Omega_-)(f^>_{p-q,k'})_n(f^>_{q-p,k})_m
\end{equation}
is the tensor of the two-photon emission corresponding to the coherent
phonon-assisted two-exciton recombination.

The causal Green's function $\hat{G}_p(\omega)$ is related to the advanced
and retarded Green's functions by the following formula\cite{23}:
\begin{equation}
\label{15}
\hat G_p(\omega)=
\frac{1}{2}\left(1+{\rm coth}\frac{\omega}{2T}\right)\hat G^R_p(\omega)+
\frac{1}{2}\left(1-{\rm coth}\frac{\omega}{2T}\right)\hat G^A_p(\omega).
\end{equation}
Substituting it in the tensor~(14) yields
\begin{eqnarray}
\label{16}
(s_p)_{nm}&=&
\frac{1}{2}\left\{\left[\left(1+{\rm coth}\frac{\Delta\omega}{2T}\right)
\hat G_{p}^R(\Delta\omega)+\left(1-{\rm coth}\frac{\Delta\omega}{2T}\right)
\hat G_p^A(\Delta\omega)\right](f^>_{-pk'})_n(f^>_{pk})_m
\right. \nonumber \\
&{+}& \left.
\left[\left(1{+}{\rm coth}\frac{\Delta\omega}{2T}\right)
\hat G_{p{-}q}^R(\Delta\omega)
{+}\left(1{-}{\rm coth}\frac{\Delta\omega}{2T}\right)
\hat G_{p-q}^A(\Delta\omega)\right](f^>_{p-q,k'})_n(f^>_{q-p,k})_m\right\},
\end{eqnarray}
where $\Delta\omega=\omega-\Omega_-$.

Using this expression, we calculate the sum over ${\bf p}$ in Eq.~(13) for
the cross section of the stimulated two-photon emission:
\begin{eqnarray}
\label{17}
&&\sum_{p\neq q/2}^{}|(s_p)_{nm}e'^*_{n}e^*_{m}|^2
\nonumber \\
&&=
\frac{1}{2}\sum_{p\neq q/2}^{}\left\{
2\left[\left(1+{\rm coth}^2\frac{\Delta\omega}{2T}\right)
|\hat G^R_p(\Delta\omega)|^2+
\left(1-{\rm coth}^2\frac{\Delta\omega}{2T}\right)
{\rm Re}\left[\hat G^R_p(\Delta\omega)\right]^2 \right]
\right.
\nonumber \\
&& \left. \times
|({\bf e'^*f}^>_{-pk'})({\bf e^*f}^>_{pk})|^2
\right.
\nonumber \\
&& \left. +
\left[\left(1+{\rm coth}\frac{\Delta\omega}{2T}\right)
\hat G^R_{p}(\Delta\omega)+
\left(1-{\rm coth}\frac{\Delta\omega}{2T}\right)
\hat G^{R*}_{p-q}(\Delta\omega)\right]
\right.
\nonumber \\
&& \left. \times
\left[\left(1+{\rm coth}\frac{\Delta\omega}{2T}\right)
\hat G^{R*}_{p-q}(\Delta\omega)+
\left(1-{\rm coth}\frac{\Delta\omega}{2T}\right)
\hat G^{R}_{p-q}(\Delta\omega)\right]
\right.
\nonumber \\
&& \left. \times
({\bf e'^*f}^>_{-pk'})({\bf e^*f}^>_{pk})
({\bf e'f}^{>*}_{p-q,k'})({\bf ef}^{>*}_{q-p,k})\right\}.
\end{eqnarray}
In deriving this formula, we have taken into account the relation between
the advanced and retarded Green's functions on the real axis of $\omega$:
$G_p^{\rm A}(\omega)=G_p^{{\rm R}*}(\omega)$.

Further calculation of the stimulated two-photon emission cross section~(13)
requires an expression for the retarded anomalous Green's function of
excitons at a finite temperature. It can be obtained through the analytical
continuation of the anomalous Green's function in the Matsubara
representation to the upper half-plane of $\omega$.

The anomalous Green's function of a Bose system in the Matsubara
representation is given by the following expression\cite{24}:
\begin{equation}
\label{18}
\hat G_p(\omega_s)=
-\frac{(1-\delta_p)\Sigma^{02}_{\omega_s p}}{
\left(i\omega_s-\epsilon_0(p)+\mu-\Sigma^{11}_{\omega_s p}\right)
\left(i\omega_s+\epsilon_0(p)-\mu+\Sigma^{11}_{-\omega_s,-p}\right)
+\Sigma^{20}_{\omega_s p}\Sigma^{02}_{\omega_s p}},
\end{equation}
where $\omega_s=2\pi sT$ and $s$ is integer. Here $\epsilon_0(p)=p^2/2m$
and $\mu$ is the system chemical potential defined by the formula
$\mu=[\sum^{11}_{\omega_sp}-\sum^{02}_{\omega_sp}]|_{\omega_s=p=0}$.

At $T\sim T_{\rm c}$, where $T_{\rm c}$ is the Bose condensation temperature
of an ideal Bose gas, the self-energy parts of a dilute Bose system
with interparticle interaction can be expressed as follows\cite{24}:
\begin{equation}
\label{19}
\Sigma^{11}_{\omega_s p}=\frac{8\pi}{m}na,\quad
\Sigma^{20}_{\omega_s p}=\Sigma^{02}_{\omega_s p}=\frac{4\pi}{m}n_0(T)a,
\end{equation}
where $n$ is the total density of particles, $a$ is the amplitude of their
mutual scattering, $n_0(T)$ is the total density of particles in the Bose
condensate, which can be approximately calculated by the formula
$n_0(T)=n[1-(T/T_{\rm c})^{3/2}]$.

Thus, the anomalous Green's function for a dilute excitonic gas can be
expressed as
\begin{equation}
\label{20}
\hat G_p(\omega_s)=(1-\delta_p)\frac{\zeta(T)}{\omega^2_s+\epsilon^2_p},
\end{equation}
where
\begin{displaymath}
\epsilon_p=\sqrt{\xi^2_p-\zeta^2(T)},\quad
\xi_p=\frac{p^2}{2m}+\zeta(T),\quad
\zeta(T)=\mu(0)\left[1-\left(\frac{T}{T_c}\right)^{3/2}\right],\quad
\mu(0)=\frac{4\pi na}{m},
\end{displaymath}
$n$ is the density of excitons and $m$ is the exciton mass. The parameter
$\mu(0)$ is the chemical potential of excitons at $T=0$.

The analytical continuation of $\hat{G}(\omega_s)$ to the upper half-plane
yields the expression for the retarded anomalous Green's function:
\begin{equation}
\label{21}
\hat G^{R}_p(\omega)=-(1-\delta_p)
\frac{\zeta(T)}{(\omega-\epsilon_p+i\Gamma_p/2)
(\omega+\epsilon_p+i\Gamma_p/2)}.
\end{equation}
Here $\Gamma_p=\tau_p^{-1}$, $\tau_p$ is the lifetime of a quasiparticle
with momentum $p$ in the excitonic system.

By substituting Eq.~(21) in (17), one can easily find that the main
contribution to the cross section of the stimulated two-photon emission~(13)
at $|\Delta\omega|\gg\Gamma_p$ is due to the summands in which
$\epsilon_p\sim|\Delta\omega|$. Therefore, matrix elements ${\bf f}_{-pk'}^>$
and ${\bf f}_{pk}^>$ can be replaced by their values
corresponding to the momentum $p_L$ that satisfies the condition
$\epsilon(p_L)=\Delta\omega$ and carried out of the integrand. Moreover,
if the $p_L\gg q$ is fulfilled, one can set $q=0$ in the sum over $p$ in
Eq.~(17). Thus, we have
\begin{eqnarray}
\label{22}
&&\sum_{\bf p}^{}|(s_p)_{nm}e'^*_{n}e^*_{m}|^2=
2\left[\left(1+{\rm coth}^2\frac{\Delta\omega}{2T}\right)
\sum_{p}^{}|\hat G^R_p(\Delta\omega)|^2
\right.
\nonumber\\
&&\quad\left.+
\left(1-{\rm coth}^2\frac{\Delta\omega}{2T}\right)
\sum_{p}^{}{\rm Re}[\hat G^R_p(\Delta\omega)]^2 \right]
|f_n(\omega'_L)f_m(\omega_L)e'^*_ne^*_m|^2,
\end{eqnarray}
where
\begin{displaymath}
{\bf f}(\omega_L)=\frac{1}{4\pi}\,\int{\bf f}^>(p_L,k)do_{p_L},\quad
{\bf f}(\omega'_L)=\frac{1}{4\pi}\int{\bf f}^>(p_L,k')do_{p_L}
\end{displaymath}
are the matrix elements averaged over the directions of vector ${\bf p}_L$.

Replacing the summation in Eq.~(22) by the integration on ${\bf p}$, we
obtain
\begin{equation}
\label{23}
\sum_{\bf p}^{}|\hat G^R_p(\Delta\omega)|^2=
\int\limits_{0}^{\infty}\frac{d^3p}{(2\pi)^3}\,
\frac{\zeta^2(T)}{|(\Delta\omega+i\Gamma_p/2)^2-\epsilon^2_p|^2}.
\end{equation}
This integral diverges as $\Gamma_p\rightarrow 0$. Replacing it as a sum of
two integrals each of which converges at $\Gamma_p\rightarrow 0$ and
replacing ${\bf p}$ integration by the integration on $t=\xi_p/\zeta(T)$
yields
\begin{equation}
\label{24}
\sum_{\bf p}^{}|\hat G^R_p(\Delta\omega)|^2=
\frac{\sqrt{2m^{3}/\zeta(T)}}{2\pi^2(\beta^2_+-\beta^2_-)}
\left[\int\limits_{1}^{\infty}\frac{dt\sqrt{t-1}}{t^2-\beta_+^2}-
\int\limits_{1}^{\infty}\frac{dt\sqrt{t-1}}{t^2-\beta_-^2}\right],
\end{equation}
where
\begin{displaymath}
\beta^2_{\pm}=(\alpha_L\pm i\gamma_L)^2+1,\quad
\alpha_L=\frac{|\Delta\omega|}{\zeta(T)},\quad
\gamma_L=\frac{\Gamma_L}{2\zeta(T)},\quad
\Gamma_L=\Gamma_{p_L}.
\end{displaymath}
Thus, in calculating the integrals on the right of the resulting equation,
one can set $\beta_{\pm}^2=\beta^2\pm i\delta$. As a result, we obtain
\begin{equation}
\label{25}
\sum_{\bf p}^{}|\hat G^R_p(\Delta\omega)|^2=
\frac{\sqrt{2m^{3}\zeta(T)(\sqrt{\alpha_L^2+1}-1)}}
{4\pi\alpha_L\Gamma_L
\sqrt{\alpha^2_L+1}}.
\end{equation}

The second sum over ${\bf p}$ in Eq.~(22) converges even as
$\Gamma_p\rightarrow 0$. Therefore, if $|\Delta\omega|\gg\Gamma_p$, we can
set in this sum $\Gamma_p=0+$. In this case, we have
\begin{equation}
\label{26}
{\rm Re}\sum_{p}^{}[\hat G^R_p(\Delta\omega)]^2=
-\frac{\sqrt{2m^{3}/\zeta(T)}}{16\pi}\,
\frac{\sqrt{\sqrt{\alpha^2_L+1}-1}\left(\sqrt{\alpha^2_L+1}+2\right)}
{\alpha_L\sqrt{(\alpha^2_L+1)^{3}}}.
\end{equation}
It is clear that for $|\Delta\omega|\gg\Gamma_p$ the following relation
takes place:
\begin{displaymath}
\left|\sum_{\bf p}^{}{\rm Re}\left[G^{R}_p(\Delta\omega)\right]^2\right|\ll
\sum_{p}^{}|G^{R}_p(\Delta\omega)|^2.
\end{displaymath}
Thus,
\begin{eqnarray}
\label{27}
&&\sum_{\bf p}^{}|(s_p)_{nm}e'^*_{n}e^*_{m}|^2=
\frac{\sqrt{2m^{3}\zeta(T)\left(\sqrt{\alpha^2_L+1}-1\right)}}{
2\pi \alpha_L\Gamma_L\sqrt{\alpha^2_L+1}}
\left(1+{\rm coth}^2\frac{\Delta\omega}{2T}\right)
\nonumber\\
&&\quad\times
|f_n(\omega'_L)f_m(\omega_L)e'^*_me^*_n|^2.
\end{eqnarray}
By substituting this expression in the formula for the differential cross
section~(13), we obtain
\begin{eqnarray}
d\sigma^L&=&
\frac{\omega(2\Omega_--\omega)^3}{4\pi c^4}\,
\frac{\sqrt{2m^{3}\zeta(T)\left(\sqrt{\alpha^2_L+1}-1
\right)}}{\alpha_L\Gamma_L\sqrt{\alpha_L^2+1}}
\left(1+{\rm coth}^2\frac{\Delta\omega}{2T}\right)
\nonumber \\
&\times&
|f_n(\omega'_L)f_m(\omega_L)e'^*_ne^*_m|^2do'.
\label{28}
\end{eqnarray}

If the exciton-phonon system is isotropic, and the incident light is
monochromatic and has a linear polarization, one has
$|e_m^*f_m(\omega_L)|^2={\bf f}^2(\omega_L)/3$. Summing over the polarizations
of photon $\omega'$ and integrating over the directions of its momentum
(note that in the case of stimulated two-photon emission the
photon $\omega$ is identical to the incident one), we obtain the total
cross section of the stimulated two-photon emission corresponding to the
coherent phonon-assisted two-exciton recombination:
\begin{equation}
\label{29}
\sigma^L(\omega,T)=\frac{\omega(2\Omega_--\omega)^3}{c^4}\,
\frac{\sqrt{8m^{3}\zeta(T)\left(\sqrt{\alpha_L^2+1}-1\right)}}{
9\alpha_L\Gamma_L\sqrt{\alpha_L^2+1}}\left(1+{\rm coth}^2\frac{\Delta\omega}
{2T}\right){\bf f}^2(\omega_L){\bf f}^2(\omega'_L).
\end{equation}

It should be noted that, if the conditions $\Delta\omega\ll\Omega_-$,
$\mu(0)\ll\Omega$, and $\tau^L={\rm const}$ are fulfilled, then, at a given
ratio between the exciton chemical
potential $\mu(0)$ at zero temperature and twice the temperature of their
Bose condensation, $\gamma=\mu(0)/2T_{\rm c}$, the parameter
$\sigma^L(\Delta\omega,T)/\sigma^L(0,0)$ is uniquely determined by two
quantities, $x=\Delta\omega/2T_{\rm c}$ and $y=T/T_{\rm c}$:
\begin{equation}
\label{30}
\frac{\sigma^L(\Delta\omega,T)}{\sigma^L(0,0)}=
\frac{z^2\sqrt{\sqrt{x^2+z^2}-z}}{|x|\sqrt{2\gamma(x^2+z^2)}}
\left(1+{\rm coth}^2\frac{x}{y}\right),
\end{equation}
where $z=\gamma(1-y^{3/2})$.

The dependence of the cross section~(29) on frequency (strictly speaking, on
the difference between the incident light frequency $\omega$ and $\Omega_-$)
is shown in Fig.~2a for different temperatures of the excitonic subsystem.
This cross section as a function of temperature at different fixed values of
the difference $\Delta\omega=\omega-\Omega_-$ is shown in Fig.~2b.
All the curves in Fig.~2 correspond to $\gamma=0.3$, and it is assumed that
$\tau^L={\rm const}$. It is clear that at $|\Delta\omega|\ll T_{\rm c}$ and
$T<T_{\rm c}$ there is a temperature interval where the cross section~(29)
of the stimulated two-photon emission is a nonmonotonic function of
temperature: $\sigma^L$ increases with the growth of temperature and can
even become larger than it is at $T=0$.

The reason for this unusual temperature dependence is the following. The
cross section~(29) of the stimulated two-photon emission is determined by two
quantities that depend on the temperature differently, namely, by $\zeta(T)$,
proportional to the number of excitons in the condensate, and by occupation
numbers of quasiparticle levels of the excitonic system with the
quasiparticle energy $\epsilon(p_L)=|\Delta\omega|$. Really, the density of
condensate and, hence, $\zeta(T)$ decreases as the temperature increases.
It leads, in turn, to a decrease in the cross
section~(29). On the other hand, using Bogoliubov's $u-v$ transforms, one
can easily show that the coherent two-exciton recombination, which is a
second-order process with respect to Hamiltonian~(3), proceeds via
intermediate states of the excitonic system containing one particle more
(less) than the state of thermodynamic equilibrium (see also Refs.~19 and
20). The cross section of the stimulated two-photon emission corresponding
to the coherent two-exciton recombination is proportional to
\begin{displaymath}
(n_{p_L}+1)^2+n^2_{p_L}=\frac{1}{2}
\left(1+{\rm coth}^2\frac{\Delta\omega}{2T}\right),
\end{displaymath}
where $n_{p_L}=[\exp(\epsilon_{p_L}/T)-1]^{-1}$ is the occupation number of
the quasiparticle state with energy $\epsilon(p_L)=|\Delta\omega|$ in the
excitonic system. As the temperature increases, $n_p$ also rises, which
leads to a larger cross section~(29). If this tendency dominates, the cross
section of the stimulated two-photon emission corresponding to the coherent
two-exciton recombination should increase with the temperature. Of course,
the tendency to decrease the cross section should overcome sooner or later as
$T\rightarrow T_{\rm c}$, since it must turn to zero at $T=T_{\rm c}$.

Note that the temperature dependence of the cross section~(29) of the
stimulated two-photon emission accompanied by the coherent two-exciton
recombination has been calculated in the approximation~(19), which is
correct only in a narrow temperature interval about the Bose condensation
temperature $T_{\rm c}$, which is considered to be equal to Bose
condensation temperature
in an ideal Bose gas. Although this approximation allows one to reproduce
formally our results\cite{20} for $T=0$, in the intermediate temperature
interval the curve of the cross section $\sigma^L(\Delta\omega,T)$ versus
temperature should be different from that plotted in Fig.~2. Nonetheless,
the conclusion about the nonmonotonic temperature dependence of the cross
section of the stimulated two-photon emission is valid. For example, at
$\Delta\omega/2T_{\rm c}=0.2$ we have
$\sigma^L(\Delta\omega,T)>\sigma^L(\Delta\omega,0)$ even for $T_{\rm
c}-T\ll T_{\rm c}$ (Fig.~2b), where the approximation~(19) is correct.

\section{Raman light scattering}

The coherent two-exciton recombination can be revealed not only in the
stimulated two-photon emission but also in the Raman light scattering
(RLS). Abrikosov and Falkovsky\cite{25} analyzed RLS in a superconductor,
whose analogue in a semiconductor was RLS by a dense electron-hole
plasma with coupling between electrons and holes (a phase transition in this
system was studied by Keldysh and Kopaev,\cite{26} see also the review by
Kopaev\cite{27} and references therein). But we are discussing the case of
a low density of electrons and holes (excitonic gas). Moreover, it is
essential for the case of RLS under consideration that the electron-hole
system is not in equilibrium, because this is the situation when the coherent
two-exciton recombination (generation) leading to a transition of an exciton
system to a state with a lower (higher) energy is possible. In the case of
such RLS with the transfer of the recoil momentum to two generated optical
phonons the energy conservation is described by the formula
\begin{equation}
\label{31}
\omega+2\Omega_-=\omega'.
\end{equation}
The case considered here corresponds to the appearance of an anti-Stokes RLS
component at frequency $\omega'$ defined by this formula.

In addition, an RLS process with the coherent two-exciton generation is also
possible, and the energy conservation in this case is described by the
equation\footnote{In a general case, RLS, like the two-photon emission, can
involve an arbitrary number of phonons. Moreover, the recoil momentum
(entirely or partially) can be transferred to impurities. Thus, RLS accompanied
by the coherent two-exciton generation or recombination can lead to the
appearance of anti-Stokes and Stokes components at frequencies
$\omega+(2\Omega-n\omega_0^s)=\omega'$ and
$\omega-(2\Omega-n\omega_0^s)=\omega'$, respectively, where $n$ is an
arbitrary integer (see also Ref.~20).}
\begin{equation}
\label{32}
\omega-2\Omega_-=\omega'.
\end{equation}
This formula determines the frequency $\omega'$ of the Stokes component
corresponding to this Raman scattering. It is clear that RLS with the coherent
phonon-assisted two-exciton generation is possible only when
$\omega>2\Omega_-$. The stimulated two-photon emission corresponding to the
coherent phonon-assisted two-exciton recombination is impossible in this
case.

The analysis of RLS accompanied by the coherent two-exciton recombination
(or generation) is similar to that of the stimulated two-photon emission
with the coherent two-exciton recombination. Since the formulas for the
cross section of RLS with the coherent two-exciton recombination or
generation are cumbersome, here we only indicate how these formulas can be
derived from Eq.~(29) using appropriate substitutions.

1.~The cross section of RLS accompanied by the coherent phonon-assisted
two-exciton recombination is obtained by replacing some variables in Eq.~(29):
\begin{displaymath}
{\bf f}(\omega_L)\rightarrow{\bf f}'(\widetilde{\omega}_L),\quad
{\bf f}(\omega'_L)\rightarrow{\bf f}(\widetilde{\omega}'_L),\quad
\omega\rightarrow-\omega,\quad
\Delta\omega\to\omega+\Omega_-,\quad
\alpha_L\rightarrow\widetilde{\alpha}_L,\quad
\Gamma_L\rightarrow\widetilde{\Gamma}_L.
\end{displaymath}
Here $\tilde{\alpha}_L=(\Omega_-+\omega)/\zeta(T)$, $\tilde{\Gamma}_L$ is
the reciprocal lifetime of a quasiparticle with energy
$\epsilon(\tilde{p}_L)=\Omega_-+\omega$ in the excitonic system,
\begin{displaymath}
{\bf f}'(\widetilde{\omega}_L)=\frac{1}{4\pi}
\int {\bf f}'^>(\widetilde{p}_L,k)do_{\tilde{p}_L},\quad
{\bf f}(\widetilde{\omega}'_L)=\frac{1}{4\pi}
\int {\bf f}^>(\widetilde{p}_L,k')do_{\tilde{p}_L}.
\end{displaymath}

2.~The cross section of RLS due to the coherent phonon-assisted two-exciton
generation ($\omega>2\Omega_-$) is derived from Eq.~(29) by substituting
\begin{displaymath}
{\bf f}(\omega_L)\rightarrow{\bf f}'(\omega_L),\quad
2\Omega_--\omega\to\omega-2\Omega_-,
\end{displaymath}
where
\begin{displaymath}
{\bf f}'(\omega_L)=\frac{1}{4\pi}\int{\bf f}'^>(p_L,k)do_{p_L}.
\end{displaymath}

\section{Possibility of experimental detection of two-photon processes
accompanied by coherent two-exciton recombination}

Let us analyze the possibility of the experimental detection of the
stimulated two-photon emission and RLS accompanied by the coherent
phonon-assisted two-exciton recombination. First we consider the stimulated
two-photon emission.

The light intensity $I^L(\omega')$ at frequency $\omega'=2\Omega_--\omega$
resulting from the stimulated two-photon emission with transfer of the recoil
momentum to generated optical phonons is given by the expression
\begin{equation}
\label{33}
I^L(\omega')=\frac{\omega'}{\omega}\sigma^L(\omega)I(\omega),
\end{equation}
where $\sigma^L(\omega)$ is the cross section of this process (Eq.~(29)),
$I(\omega)$ (W/cm$^2$) is the intensity of incident light of frequency
$\omega$.

The intensity~(33) can be expressed as a sum of two terms:
\begin{equation}
\label{34}
I^L(\omega')=\Delta I^L(\omega')+\widetilde{I}^L(\omega'),
\end{equation}
where $\widetilde{I}^L(\omega')$ is the intensity of the two-photon emission
resulting from two consecutive processes: the spontaneous emission at
frequency $\omega'=2\Omega_--\omega$ and the subsequent stimulated emission at
frequency $\omega$, each of which satisfies the energy conservation law.

If the incident light frequency $\omega>\Omega_-$, then $\omega'<\Omega_-$.
In this case, the spontaneous emission at frequency
$\omega'=2\Omega_--\omega<\Omega_-$is due to the excitonic recombination
with generation of a Bogoliubov quasiparticle with momentum ${\bf p}_L$
that satisfies the condition $\epsilon(p_L)=\Delta\omega$ (see Appendix B,
$\Delta\omega=-\Delta\omega'$). The spontaneous recombination of excitons
generates in the excitonic system $I_s^L(\omega')/\omega'$ quasiparticles
with energy $\epsilon(p_L)=\Delta\omega$ per unit time, where
$I_s^L(\omega')$ is the luminescence intensity~(57) (see Appendix B and
Ref.~28). These quasiparticles disappear in the time of order of $\tau^L$.
The disappearance of some of these quasiparticles is accompanied by the
stimulated recombination of excitons and the induced emission of light at
frequency $\omega$. Thus, at $\omega>\Omega_-$, the intensity
$\tilde{I}(\omega')$ is given by the relation
\begin{equation}
\label{35}
\widetilde{I}^L(\omega')=\frac{\tau^L}{\tau^L_r}I^L_s(\omega'),
\end{equation}
where $\tau_r^L$ is the lifetime of the quasiparticle with energy
$\epsilon(p_L)$ due to its recombination, which leads to the stimulated
emission at frequency $\omega$, provided that the excitonic system contains
one quasiparticle with momentum ${\bf p}_L$ more than it does in the state of
thermodynamic equilibrium. The time $\tau_r^L$ can be easily calculated
using Fermi's `golden rule':
\begin{displaymath}
\frac{1}{\tau^L_r}=\frac{(2\pi)^2}{3c}{\bf f}^2(\omega_L)u^2_{p_L}
(n_{p_L}+1)I(\omega),
\end{displaymath}

\vspace{-10mm}

\begin{equation}
\label{36}
\end{equation}

\vspace{-10mm}

\begin{displaymath}
u^2_{p_L}=\frac{1}{2}\left(\frac{\sqrt{\alpha^2_L+1}}{\alpha_L}+1\right),
\quad
n_{p_L}=\frac{1}{e^{\Delta\omega/T}-1},
\end{displaymath}
where $u_{p_L}$ is Bogoliubov's coefficient and $n_{p_L}$ is the distribution
function of quasiparticles with energy $\epsilon(p_L)=\Delta\omega$ at
temperature $T$.

If the incident light frequency $\omega<\Omega_-$ and, hence,
$\omega'>\Omega_-$, the situation is similar to that discussed above. In
this case, the spontaneous emission at frequency $\omega'$ is due to the
recombination of an exciton accompanied by the disappearance of one Bogoliubov
quasiparticle with energy $\epsilon(p_L)=-\Delta\omega$ in the excitonic
system. At $\omega'>\Omega_-$, the number of quasiparticles of energy
$\epsilon(p_L)=-\Delta\omega$ that disappear per unit time as a result
of spontaneous recombination of excitons is $I_s^L(\omega')/\omega'$.
In the time of order of $\tau^L$,
the disappeared quasiparticles are replaced by new ones, and the appearance
of some of them is accompanied by the stimulated radiation at frequency
$\omega$. Thus, at $\omega<\Omega_-$ we have
\begin{equation}
\label{37}
\widetilde{I}^L(\omega')=\frac{\tau^L}{\tau^L_c}I^L_s(\omega'),
\end{equation}
where $\tau_c^L$ is the lifetime of an exciton with momentum ${\bf p}_L$ with
respect to the stimulated recombination, which results in both a stimulated
emission at frequency $\omega$ and generation of a quasiparticle of energy
$\epsilon(p_L)=-\Delta\omega$, provided that the excitonic system contains
one quasiparticle with momentum ${\bf p}_L$ less than it does in the state of
thermodynamic equilibrium. Using Fermi's `golden rule' yields the following
expression for $\tau_c^L$:
\begin{displaymath}
\frac{1}{\tau^L_c}=\frac{(2\pi)^2}{3c}{\bf f}^2(\omega_L)
v^2_{p_L}n_{p_L}I(\omega),
\end{displaymath}

\vspace{-6.5mm}

\begin{equation}
\label{38}
\end{equation}

\vspace{-6.5mm}

\begin{displaymath}
v^2_{p_L}=\frac{1}{2}\left(\frac{\sqrt{\alpha^2_L+1}}{\alpha_L}-1\right).
\end{displaymath}
Using Eqs.~(35)-(38) and (57) from Appendix B, we obtain the intensity
$\tilde{I}^L(\omega')$ in a general case:
\begin{eqnarray}
\widetilde{I}^L(2\Omega_--\omega) &=&
\frac{2\Omega_--\omega}{\omega}
\widetilde\sigma^L(\omega)I(\omega),
\nonumber \\
\widetilde{\sigma}^L(\omega) &=&
\tau^L\frac{\omega(2\Omega_--\omega)^3}{c^4}\,
\frac{\sqrt{2m^{3}\zeta(T)\left(\sqrt{\alpha^2_L+1}-1
\right)}}{18\alpha_L\sqrt{\alpha^2_L+1}}
{\bf f}^2(\omega_L){\bf f}^2(\omega'_L)
\nonumber \\
&&\quad
\times \left[{\rm sign}(\Delta\omega)+{\rm coth}\frac{|\Delta\omega|}{2T}
\right]^2.
\label{39}
\end{eqnarray}

The spectral line at frequency $\omega'=2\Omega_--\omega$ due to the
stimulated two-photon emission accompanied by the coherent phonon-assisted
two-exciton recombination will be observed against the background of the
luminescence spectrum of Bose-condensed excitons. It clearly follows from
Eqs.~(35) and (37) that $\tilde{I}^L(\omega')$ determines a fraction of the
spontaneous emission intensity $I_s^L(\omega')$. Thus, the total intensity
of the emission at frequency $\omega'$ can be expressed as follows:
\begin{equation}
\label{40}
I^L_{\rm tot}(\omega')=\Delta I^L(\omega')+I^L_s(\omega'),
\end{equation}
where $I_s^L(\omega')$ is the luminescence intensity at frequency $\omega'$
in the absence of the incident light of frequency $\omega$, $\Delta
I^L(\omega')$ is the observed light intensity at frequency $\omega'$ due to
the stimulated two-photon emission with the coherent two-exciton
recombination. By substituting the cross section~(29) in Eq.~(33) and using
Eqs.~(34) and (39), we obtain the observed light intensity $\Delta
I^L(\omega')$:
\begin{eqnarray}
\Delta I^L(2\Omega_--\omega)
&=& \frac{2\Omega_--\omega}{\omega}
\Delta\sigma^L(\omega)I(\omega),
\nonumber \\
\Delta\sigma^L(\omega) &=&
\tau^L\frac{\omega(2\Omega_--\omega)^3}{c^4}\,
\frac{\sqrt{8m^{3}\zeta(T)\left(\sqrt{\alpha^2_L+1}-1\right)}}
{9\alpha_L\sqrt{\alpha^2_L+1}}{\bf f}^2(\omega_L){\bf f}^2(\omega'_L)
\nonumber \\
&&\quad
\times \left[1+{\rm coth}^2\frac{\Delta\omega}{2T}-
\frac{1}{4}\left({\rm sign}(\Delta\omega)+
{\rm coth}\frac{|\Delta\omega|}{2T}\right)^2\right].
\label{41}
\end{eqnarray}
One can easily prove that $1/2\leq\Delta\sigma^L(\omega)/\sigma^L\leq 1$.
In particular, at $T=0$ we have $\Delta\sigma^L(\omega)=\sigma^L(\omega)$
at $\omega<\Omega_-$ and $\Delta\sigma^L(\omega)=\sigma^L(\omega)/2$ at
$\omega>\Omega_-$.

Using Eq.~(29), we can estimate the cross section $\sigma^L$ of the stimulated
two-photon emission. In CGS units this expression has the form
\begin{eqnarray}
\sigma^L(\omega,T)&=&\tau^LV
\frac{\omega(2\Omega_--\omega)^3}{9c^4\hbar^4}\,
\frac{\sqrt{8m^{3}\zeta(T)\left(\sqrt{\alpha_L^2+1}-1
\right)}}{\alpha_L\sqrt{\alpha_L^2+1}}
\nonumber \\
&\times&
\left(1+{\rm coth}^2\frac{\Delta\omega}{2T}\right)
{\bf f}^2(\omega_L){\bf f}^2(\omega'_L),
\label{42}
\end{eqnarray}
where $V$ is the volume of the excitonic system exposed to the incident light
and $\alpha_L=\hbar|\Delta\omega|/\zeta(T)$.

We shall consider as an example a system of Bose-condensed excitons in
Cu$_2$O at zero temperature. The exciton effective mass in this crystal is
$m=2.7m_e$, the characteristic exciton radius is equal to $a=7$~\AA, and the
photon energy corresponding to the recombination of an exciton with zero
momentum is $\hbar\Omega\simeq 2$~eV. The optical recombination of an exciton
in Cu$_2$O is typically assisted by generation of an optical phonon of energy
$\hbar\omega_0^s\simeq 10$~meV with a negligible dispersion.

Let us estimate the exciton chemical potential at $T=0$ by the formula
\begin{eqnarray*}
\mu(0)=\frac{4\pi\hbar^2}{m}na,
\end{eqnarray*}
where $n$ is the exciton density. Assuming that $n=10^{19}$~cm$^{-3}$ (this
density was achieved in some experiments\cite{9}), we obtain $\mu(0)\simeq
2.5$~meV. An ideal gas of excitons with $n=10^{19}$~cm$^{-3}$ should
transform to the Bose-condensed state at $T_{\rm c}\sim 50$~K, in this case
$\mu(0)/2T_{\rm c}\simeq 0.3$.

In the experiment\cite{9} excitons were generated by powerful nanosecond
laser pulses at a wavelength $\lambda\simeq 500$~nm focused into the spot of
diameter $d\simeq 30\;\mu$m on the sample surface. The volume of the
excitonic system interacting with the incident light stimulating the
two-photon emission can be estimated as $V=d^2l$, where $l\simeq 1\;\mu$m is
the penetration depth of radiation with wavelength 500~nm.

As $\omega\rightarrow\Omega_-$ ($\alpha_L\rightarrow 0$), the cross section
$\sigma^L$ increases. Let us suppose that $\hbar(\Omega_--\omega)=\mu(0)$. In
this case
\begin{displaymath}
{\bf f}(\omega_L)\simeq{\bf f}(\omega'_L)\simeq {\bf F},\quad
{\bf F}=\frac{1}{4\pi}\int {\bf F}^>(p_L,k)do_{p_L},
\end{displaymath}
where ${\bf F}$ is the matrix element of the radiative phonon-assisted
recombination of an isolated exciton.\cite{20} This matrix element can be
estimated by the formula
\begin{equation}
\label{43}
\frac{1}{\tau_{\rm exc}}=\frac{4\Omega^3_-}{3c^3\hbar}{\bf F}^2,
\end{equation}
where $\tau_{\rm exc}$ is the lifetime of an isolated exciton due to its
spontaneous recombination with the emission of a photon of energy
$\hbar\Omega_-$ and an optical phonon with energy $\hbar\omega_0^s$. The
lifetime of para-excitons in Cu$_2$O is $\tau_{\rm exc}\sim
100\;\mu$s.\cite{7}

The relaxation time $\tau^L$ in the system of Bose-condensed excitons is a
subject of further investigation. Even at zero temperature, it can be
considerably shorter than the radiative lifetime of excitons $\tau_{\rm
exc}$, because a quasiparticle can disappear, for example, due to the emission
of one or several acoustic phonons. Assuming that the time $\tau^L$ is
within the interval of $10^{-11}$--$10^{-5}$~s (the lower bound is defined
by the condition $\Gamma_L=10^{-1}\epsilon(p_L)$, the upper bound is
$10^{-1}\tau_{\rm exc}$), we obtain an estimate for the cross
section of the stimulated two-photon emission by Bose-condensed para-excitons
in Cu$_2$O at $T=0$: $\sigma^L=10^{-16}$--$10^{-10}$~cm$^2$.

The radiative lifetime of ortho-excitons in Cu$_2$O is $\tau_{\rm
exc}\sim 300$~ns. Assuming the relaxation time $\tau^L$ in a system of
ortho-excitons in the Bose-condensed state is within
$10^{-11}$--$10^{-9}$~s (in this case the upper bound is determined by the
time of transition between the ortho-exciton and para-exciton states), yields
$\sigma^L=10^{-11}$--$10^{-9}$~cm$^2$ at $T=0$. Thus, the stimulated
two-photon emission accompanied by the coherent two-exciton recombination
can be experimentally detected in Cu$_2$O.

The cross section of RLS with the coherent two-exciton recombination
accompanied by the generation of two optical phonons is determined by the
squared product of two matrix elements:
\begin{displaymath}
{\bf f}'(\widetilde{\omega}_L)=\frac{1}{4\pi}\,\int{\bf
f}'^>(\widetilde{p}_L,k) do_{\tilde{p}_L},\quad
{\bf f}(\widetilde{\omega}'_L)=\frac{1}{4\pi}
\int {\bf f}^>(\widetilde{p}_L,k')do_{\tilde{p}_L},
\end{displaymath}
where $\tilde{p}_L$ is determined by the condition
$\epsilon(\tilde{p}_L)=\omega +\Omega_-$ (see Section~3). The band gap in
Cu$_2$O is wide ($\Omega_-\sim 10^2\omega_0^s$), therefore
$\epsilon(\tilde{p}_L)\gg \omega_0^s$. Using the approach
suggested in Appendix A, one can prove that in this case ${\bf
f}(\tilde{\omega}'_L)$ and ${\bf f'}(\tilde{\omega}_L)$ are negligible in
comparison with the matrix elements ${\bf f}(\omega_L)$ and ${\bf
f}(\omega'_L)$ in Eq.~(42) at $|\Omega_--\omega|\sim\mu(0)$. Moreover, the
cross section of the RLS under consideration is proportional to the lifetime
of a quasiparticle with energy $\epsilon(\tilde{p}_L)=\omega+\Omega_-$,
which is essentially shorter than the relaxation time $\tau^L$ in the cross
section~(42) at $|\Omega_--\omega|\sim\mu(0)$. Thus, unlike the stimulated
two-photon emission, one can hardly detect RLS accompanied by the coherent
two-exciton recombination in Cu$_2$O. The situation is similar in the case of
RLS with the coherent two-exciton generation.

\section{Conclusions}

In this paper, we have demonstrated that the coherent two-exciton
recombination, i.e., the simultaneous recombination of two
excitons with opposite momenta corresponding to the existence of
nondiagonal long-range order in the system expressed by nonvanishing
anomalous averages of the form $\langle N-2|Q_{-p}Q_p|N\rangle$, is possible
in Bose-condensed excitonic system interacting with the electromagnetic field.
Similarly, the coherent two-exciton generation corresponding to anomalous
averages like $\langle N-2|Q^+_{-p}Q^+_p|N\rangle$ is also possible. In these
processes, the exciton occupation numbers are unchanged, and the final and
state of the excitonic system differs from the initial one only in the
average number of excitons with zero momentum. The coherent two-exciton
recombination may also lead to Raman light scattering by excitons in
Bose-condensed state (RLS can also be accompanied by the coherent two-exciton
generation). The recoil momentum corresponding to the change in the momentum
of electromagnetic field is transferred to phonons or impurities. Both the
stimulated two-photon emission and RLS with the coherent two-exciton
recombination (generation) can occur only in the presence of Bose condensate
in a system of interacting excitons, therefore, the observation of these
effects can be used as a strong experimental evidence of the existence of
Bose condensation in excitonic systems.

Using the diagrammatic methods, we have developed a technique for
calculating the cross sections of the stimulated two-photon emission and
RLS accompanied by the coherent two-exciton recombination (or generation) at
$T>0$. In this approach, the elements of the scattering matrix corresponding
to the processes in question are expressed in a natural manner in terms of
Green's functions of Bose-condensed excitons (see Eqs.~(9), (10), and also
(49)).

If the incident light frequency $\omega<2\Omega_-$, where
$\Omega_-=\Omega-\omega_0^s$ ($\Omega$ is the frequency of light due to the
recombination of an exciton with zero momentum and $\omega_0^s$ is the
optical phonon frequency), the stimulated two-photon emission and RLS
accompanied by the coherent phonon-assisted two-exciton recombination
result in the appearance of a spectral line at frequency $2\Omega_--\omega$
and the anti-Stokes component at $\omega+2\Omega_-$, respectively. At
$\omega>2\Omega_-$ the RLS spectrum contains both the anti-Stokes and
Stokes components at frequencies $\omega\pm 2\Omega_-$. The anti-Stokes
line corresponds due to the coherent two-exciton recombination, whereas the
Stokes component is due to the coherent phonon-assisted two-exciton
generation. In this case, the stimulated two-photon emission is impossible.

Using approximation~(19), we have derived expressions for the cross
sections of the processes under consideration at finite temperatures. If
$|\omega-\Omega_-|\ll T_{\rm c}$ ($T_{\rm c}$ is the temperature of Bose
condensation), the cross section of the stimulated two-photon emission is a
nonmonotonic function of temperature. It increases in a certain temperature
interval below $T_{\rm c}$ and can even be larger than it is at $T=0$. The
cause of this nonmonotonic behavior is that the cross section of the
stimulated two-photon emission is determined not only by the density of
excitons in the condensate, which decreases as the temperature increases and
vanishes at $T=T_{\rm c}$, but also by the occupation numbers of
quasiparticles with energies $|\omega-\Omega_-|$ in the excitonic system,
which increases as the temperature grows.

Our estimates indicate that, at $|\omega-\Omega_-|\sim\mu(0)$, where
$\mu(0)$ is the exciton chemical potential measured with respect to the
excitonic band bottom, a spectral line at $2\Omega_--\omega$ due to the
stimulated two-photon emission accompanied by the coherent optical
phonon-assisted two-exciton recombination can be experimentally observed in
Cu$_2$O.

The work was supported by grants from INTAS, Russian Foundation for Basic
Research, and Physics of Solid-State Nanostructures program.

\rightline{\bf Appendix A}

\section*{Effective matrix elements of excitonic recombination}

The objective of this Appendix is to prove that the two-photon emission and
RLS accompanied by the coherent two-exciton recombination can be analyzed on
the base of first principles, without using the effective Hamiltonian~(3).
Taking as a example the two-photon emission, we will determine conditions
when the analysis based on the effective Hamiltonian~(3) is correct. In
addition, we will show that the effective matrix elements of excitonic
recombination used in this paper do not depend on temperature and equal to
those calculated previously for $T=0$.\cite{20}

The Hamiltonian describing the interaction among excitons, phonons, and
electromagnetic field can be written as
\begin{eqnarray}
\hat V(t) &=& \hat W(t)+\hat D(t),
\nonumber\\
\hat W(t) &=& \sum_{pq}^{}\left[W_{qp}Q^+_q(t)Q_p(t)b_{q-p}(t)+
W^*_{pq}Q^+_q(t)Q_p(t)b^+_{p-q}(t)\right],
\label{44}\\
\hat D(t) &=& \sum_{q}^{}\left[D_qe^{-i\Omega t}Q_q(t)c^+_q(t)+
D'_qe^{-i\Omega t}Q_{-q}(t)c_q(t)+{\rm H.c.}\right],
\nonumber
\end{eqnarray}
where Hamiltonian $\hat{W}(t)$ describes the scattering of excitons by
phonons, $\hat{D}(t)$ is responsible for the interaction between excitons and
the electromagnetic field, $D_q=i\sqrt{2\pi\omega_q}{\bf e^*d}_q$,
$D'_q=-i\sqrt{2\pi\omega_q}{\bf ed}_q$.

It is clear that the two-photon emission accompanied by the coherent
phonon-assisted two-exciton recombination is a process of the fourth order
in Hamiltonian $\hat{V}(t)$. For the element of the $S$-matrix of the
two-photon emission due to the coherent two-exciton recombination and
generation of two optical phonons with momenta ${\bf p-k}$ and ${\bf
-p-k'}$ averaged with the Gibbs distribution, we have
\begin{eqnarray}
(S_p)_{fi}&=&\frac{(-i)^4}{4!}\int\limits_{-\infty}^{\infty}
\ldots
\int
\langle f|T_t\left[\hat W(t_1)\hat W(t_2)\hat D(t_3)\hat D(t_4)
\right.
\nonumber \\
&+& \left.
\hat W(t_1)\hat D(t_2)\hat W(t_3)\hat D(t_4)+
\hat D(t_1)\hat W(t_2)\hat W(t_3)\hat D(t_4)+
\hat W(t_1)\hat D(t_2)\hat D(t_3)\hat W(t_4)
\right.
\nonumber\\
&+& \left.
\hat D(t_1)\hat W(t_2)\hat D(t_3)\hat W(t_4)+
\hat D(t_1)\hat D(t_2)\hat W(t_3)\hat W(t_4)\right]|i\rangle dt_1...dt_4.
\label{45}
\end{eqnarray}
Here $\langle f|\ldots|i\rangle=\sum\limits_n\exp[(F-E_n(N)+\mu
N)/T]\langle m|\ldots|n\rangle$, where $|n\rangle=|n,N\rangle_{\rm
exc}|i\rangle_{\rm phon}|i\rangle_{\rm phot}$ and
$|m\rangle=|n,N-2\rangle_{\rm exc}|f\rangle_{\rm phon}|f\rangle_{\rm
phot}$, and the other notations are given in Section~2.

By changing the time variables in each summand of Eq.~(45) we can transform
it to
\begin{equation}
\label{46}
(S_p)_{fi}=\frac{1}{4}\int\limits_{-\infty}^{\infty}
\ldots
\int
\langle f|T_t\hat W(t_1)\hat W(t_2)\hat D(t_3)\hat D(t_4)|i\rangle
dt_1\ldots dt_4,
\end{equation}
where
\begin{eqnarray}
&&\langle f|T_t\hat W(t_1)\hat W(t_2)\hat D(t_3)\hat D(t_4)|i\rangle=
D_kD_{k'}\exp\left[{-i\Omega(t_3+t_4)}\right]
\sum_{p_1p_2}^{}W_{p_1,p_1+p+k'}W_{p_2,p_2-p+k}
\nonumber \\
&&\quad\times
\left\{\left[\langle T_tQ^+_{p_1+p+k'}(t_1)Q_{p_1}(t_1)Q^+_{p_2-p+k }
(t_2)Q_{p_2}(t_2)Q_{k}(t_3)Q_{k'}(t_4)\rangle\times
\right.\right.
\nonumber\\
&&\quad\times\left.\left.
\langle f|T_tb^+_{-p-k'}(t_1)b^+_{p-k}(t_2)|i\rangle_{\rm phon}
\right.\right.
\nonumber\\
&&\quad+\left.\left.
\langle T_tQ^+_{p_2-p+k}(t_1)Q_{p_2}(t_1)Q^+_{p_1+p+k'}(t_2)Q_{p_1}(t_2)
Q_{k}(t_3)Q_{k'}(t_4)\rangle
\right.\right.
\nonumber\\
&&\quad\times\left.\left.
\langle f|T_tb^+_{p-k}(t_1)b^+_{-p-k'}(t_2)|i\rangle_{\rm phon}\right]
\langle f|T_tc^+_k(t_3)c^+_{k'}(t_4)|i \rangle_{\rm phot}
\right.
\nonumber\\
&&\quad+\left.\left[
\langle T_tQ^+_{p_1+p+k'}(t_1)Q_{p_1}(t_1)Q^+_{p_2-p+k}(t_2)Q_{p_2}(t_2)
Q_{k'}(t_3)Q_{k}(t_4)\rangle
\right.\right.
\nonumber\\
&&\quad\left.\left.\times
\langle f|T_tb^+_{-p-k'}(t_1)b^+_{p-k}(t_2)|i\rangle_{\rm phon}
\right.\right.
\nonumber\\
&&\quad\left.\left.+
\langle T_tQ^+_{p_2-p+k}(t_1)Q_{p_2}(t_1)Q^+_{p_1+p+k'}(t_2)Q_{p_1}(t_2)
Q_{k'}(t_3)Q_{k}(t_4)\rangle
\right.\right.
\nonumber\\
&&\quad\left.\left.\times
\langle f|T_tb^+_{p-k}(t_1)b^+_{-p-k'}(t_2)|i\rangle_{\rm phon}
\right]
\langle f|T_tc^+_{k'}(t_3)c^+_{k}(t_4)|i\rangle_{\rm phot}\right\}.
\label{47}
\end{eqnarray}
Here $\langle\ldots\rangle=\sum\limits_n\exp[(F-E_n(N)+\mu
N)/T]\langle n,N-2|\ldots|n,N\rangle_{\rm exc}$.

In the most interesting case $k\ne k'$, we have
\begin{eqnarray}
&&\sum_{p_1p_2}^{}W^*_{p_1,p_1+p+k'}W^*_{p_2,p_2-p+k}
\langle T_tQ^+_{p_1+p+k'}(t_1)Q_{p_1}(t_1)Q^+_{p_2-p+k}(t_2)
Q_{p_2}(t_2)Q_{k}(t_3)Q_{k'}(t_4)\rangle
\nonumber\\
&&\quad=
\sum_{p_1}^{}W^*_{p_1 p_1}\left[W^*_{-k'k}
\langle T_t Q^+_{p_1}(t_1)Q_{p_1}(t_1)\rangle
\langle T_t Q^+_{k}(t_2)Q_{k}(t_3)\rangle
\langle T_t Q_{-k'}(t_2)Q_{k'}(t_4)\rangle
\right.
\nonumber\\
&&\quad+\left.
W^*_{-kk'}\langle T_t Q^+_{p_1}(t_1)Q_{p_1}(t_1)\rangle
\langle T_t Q^+_{k'}(t_2)Q_{k'}(t_4)\rangle
\langle T_t Q_{-k}(t_2)Q_{k}(t_3)\rangle\right]\delta(p+k')
\nonumber\\
&&\quad+
\sum_{p_2}^{}W^*_{p_2 p_2}\left[W^*_{-k'k}
\langle T_t Q^+_{-k} (t_1)Q_{k}(t_3)\rangle
\langle T_t Q_{-k'}(t_1)Q_{k'}(t_4)\rangle
\langle T_t Q_{p_2}(t_2)Q^+_{p_2}(t_2)\rangle
\right.
\nonumber\\
&&\quad+\left.
W^*_{-kk'}\langle T_t Q^+_{k'}(t_1)Q_{k'}(t_4)\rangle
\langle T_t Q_{-k}(t_1)Q_{k}(t_3)\rangle
\langle T_t Q^+_{p_2}(t_2)Q_{p_2}(t_2)\rangle\right]\delta(p-k)
\nonumber\\
&&\quad+
W^*_{-k,p-q}W^*_{-k',q-p}\langle T_t Q^+_{p-q}(t_1)Q^+_{q-p}(t_2)\rangle
\langle T_t Q_{-k }(t_1)Q_{k}(t_3)\rangle
\langle T_t Q_{-k'}(t_2)Q_{k'}(t_4)\rangle
\nonumber \\
&&\quad+W^*_{-k'p}W^*_{-k,-p}
\langle T_t Q^+_{p}(t_1)Q^+_{-p}(t_2)\rangle
\langle T_t Q_{-k'}(t_1)Q_{k'}(t_4)\rangle
\langle T_t Q_{-k }(t_2)Q_{k }(t_3)\rangle
\nonumber\\
&&\quad+
W^*_{q-p,k}W^*_{p-q,k'}
\langle T_t Q^+_{k}(t_1)Q_{k}(t_3)\rangle
\langle T_t Q_{q-p}(t_1)Q_{p-q}(t_2)\rangle
\langle T_t Q^+_{k'}(t_2)Q_{k'}(t_4)\rangle
\nonumber\\
&&\quad+
W^*_{-pk'}W^*_{pk}
\langle T_t Q^+_{k'}(t_1)Q_{k'}(t_4)\rangle
\langle T_t Q_{-p}(t_1)Q_{p}(t_2)\rangle
\langle T_t Q^+_{k}(t_2)Q_{k}(t_3)\rangle
\nonumber\\
&&\quad+
W^*_{-k,p-q}W^*_{p-q,k'}
\langle T_t Q^+_{p-q}(t_1)Q_{p-q}(t_2)\rangle
\langle T_t Q_{-k }(t_1)Q_{k}(t_3)\rangle
\langle T_t Q^+_{k'}(t_2)Q_{k'}(t_4)\rangle
\nonumber\\
&&\quad+
W^*_{-k'p}W^*_{pk}
\langle T_t Q^+_{p}(t_1)Q_{p}(t_2)\rangle
\langle T_t Q_{-k'}(t_1)Q_{k'}(t_4)\rangle
\langle T_t Q^+_{k}(t_2)Q_{k}(t_3)\rangle
\nonumber\\
&&\quad+
W^*_{q-p,k}W^*_{-k',q-p}
\langle T_t Q^+_{k}(t_1)Q_{k}(t_3)\rangle
\langle T_t Q_{q-p}(t_1)Q^+_{q-p}(t_2)\rangle
\langle T_t Q_{-k'}(t_2)Q_{k'} (t_4)\rangle
\nonumber\\
&&\quad+
W^*_{-pk'}W^*_{-k,-p}
\langle T_t Q^+_{k'}(t_1)Q_{k'}(t_4)\rangle
\langle T_t Q_{-p}(t_1)Q^+_{-p}(t_2)\rangle
\langle T_t Q_{-k}(t_2)Q_{k}(t_3)\rangle.
\label{48}
\end{eqnarray}
Similar expressions can be derived from the rest of the summands in
Eq.~(47).

By substituting Eq.~(48) in (46) and performing integration over
time variables, we obtain for ${\bf p+k'}\ne 0$ and ${\bf p-k}\ne 0$
(the phonons are supposed to be optical)
\begin{eqnarray}
(S_p)_{fi} &=& 2\pi iT_{k'k}({\bf p})
\left[(\sqrt{2}-1)\delta({\bf p}-{\bf q}/2)+1\right]
\delta(\omega'+\omega-2\Omega_-),
\nonumber\\
T_{k'k}({\bf p}) &=& D_{k}D_{k'}\left[W^*_{q-p,k}W^*_{p-q,k'}
G_{k }(\omega-\Omega)G_{k'}(\Omega_--\omega-\omega^s_0)
\widetilde G_{p-q}(\omega -\Omega_- )\right.
\nonumber\\
&&\quad+\left.
W^*_{-k,p-q}W^*_{-k',q-p}\widetilde G_{-k}(\omega-\Omega)
\widetilde G_{k' }(\omega-\Omega_-+\omega^s_0)
\widetilde G^+_{q-p}(\omega-\Omega_-)\right.
\nonumber\\
&& \quad+\left.
W^*_{q-p,k}W^*_{-k',q-p}G_{k }(\omega-\Omega)
\widetilde G_{k'}(\omega-\Omega_-+\omega^s_0)G_{q-p}(\omega -\Omega_-)
\right.
\nonumber\\
&&\quad+\left.
W^*_{-k,p-q}W^*_{p-q,k'}
\widetilde G_{-k}(\omega-\Omega)G_{k'}(\Omega_--\omega-\omega^s_0)
G_{p-q}(\Omega_--\omega)\right.
\nonumber\\
&&\quad+\left.
W^*_{-pk'}W^*_{pk}G_{k'}(\Omega_--\omega-\omega^s_0)G_{k}(\omega-\Omega)
\widetilde G_{-p }(\omega-\Omega_-)\right.
\nonumber\\
&&\quad+\left.
W^*_{-k'p}W^*_{-k,-p}
\widetilde G_{-k}(\omega-\Omega)\widetilde G_{k'}(\omega-\Omega_-+\omega^s_0)
\widetilde G^+_{p}(\omega-\Omega_-)\right.
\nonumber\\
&&\quad+\left.
W^*_{-k'p}W^*_{pk}G_{k }(\omega-\Omega)
\widetilde G_{k'}(\omega-\Omega_-+\omega^s_0)G_{p  }(\omega-\Omega_-)
\right.
\nonumber\\
&&\quad+\left.
W^*_{-pk'}W^*_{-k,-p }
\widetilde G_{-k}(\omega-\Omega)G_{k' }(\Omega_--\omega-\omega^s_0)
G_{-p }(\Omega_--\omega)\right],
\label{49}
\end{eqnarray}
where $\tilde{G}_p(\omega)$, $\tilde{G}_p^+(\omega)$, and $G_p(\omega)$ are
Fourier transforms of the anomalous and normal Green's functions of
excitons in the Bose-condensed state, which are defined as follows:
\begin{displaymath}
G_p(t-t')=-i\langle T_t Q_p(t)Q^+_p(t')\rangle,
\end{displaymath}

\vspace{-9mm}

\begin{equation}
\label{50}
\end{equation}

\vspace{-9mm}

\begin{displaymath}
\widetilde G^{(+)}_p(t-t')=-i\langle T_t Q^{(+)}_{-p}(t)Q^{(+)}_p(t')\rangle.
\end{displaymath}

Thus, we have derived the expression for the element of the $S$-matrix
responsible for the two-photon emission due to coherent recombination
directly from the Hamiltonian of the interaction between excitons and
the electromagnetic field and the Hamiltonian of the exciton-phonon
interaction~(44). Similarly elements of the $S$-matrix corresponding to the
RLS accompanied by the coherent two-exciton recombination (or generation) can
be derived.

In a general case, Eq.~(49) cannot be reduced to the corresponding
expression~(12) derived from the effective Hamiltonian~(3). Below we will
determine the conditions when this is possible and derive an expression for
$L_{pq}^>$.

By analyzing the stimulated two-photon emission under the condition
$|\omega-\Omega_-|\ll\omega_0^s$, we obtain
\begin{eqnarray}
T_{k'k}({\bf p})&=&
D_{k}D_{k'}\left[W^*_{q-p,k}W^*_{p-q,k'}G_{k }(-\omega^s_0)G_{k' }(-\omega^s_0)
\widetilde G_{p-q}(\omega -\Omega_-)\right.
\nonumber\\
&+&\left.
W^*_{-k,p-q}W^*_{-k',q-p}\widetilde G_{-k}(-\omega^s_0)
\widetilde G_{k' }(\omega^s_0)\widetilde G^+_{q-p}(\omega-\Omega_-)
\right.
\nonumber\\
&+&\left.
W^*_{q-p,k}W^*_{-k',q-p}G_{k }(-\omega^s_0)\widetilde G_{k' }(\omega^s_0)
G_{q-p}(\omega -\Omega_-)\right.
\nonumber\\
&+&\left.
W^*_{-k,p-q}W^*_{p-q,k'}\widetilde G_{-k}(-\omega^s_0)G_{k' }(-\omega^s_0)
G_{p-q}(\Omega_--\omega)\right.
\nonumber\\
&+&\left.
W^*_{-pk'}W^*_{pk}G_{k'}(-\omega^s_0)G_{k}(-\omega^s_0)
\widetilde G_{-p }(\omega-\Omega_-)\right.
\nonumber\\
&+&\left.
W^*_{-k'p}W^*_{-k,-p}\widetilde G_{-k}(-\omega^s_0)
\widetilde G_{k' }(\omega^s_0)\widetilde G^+_{p}(\omega-\Omega_-)
\right.
\nonumber\\
&+&\left.
W^*_{-k'p}W^*_{pk}G_{k }(-\omega^s_0)\widetilde G_{k'}(\omega^s_0)
G_{p}(\omega-\Omega_-)\right.
\nonumber\\
&+&\left.
W^*_{-pk'}W^*_{-k,-p}\widetilde G_{-k}(-\omega^s_0)G_{k' }(-\omega^s_0)
G_{-p }(\Omega_--\omega)\right].
\label{51}
\end{eqnarray}

Under the conditions of approximation~(19), we have for the retarded
Green's functions
\begin{displaymath}
G^R_p(\omega)=-2\pi in_0(T)\delta_p\delta(\omega)+G'^R_p(\omega),
\end{displaymath}

\vspace{-8mm}

\begin{equation}
\label{52}
\end{equation}

\vspace{-8mm}

\begin{displaymath}
G'^{R}_p(\omega)=(1-\delta_p)\frac{\omega+\xi_p}{(\omega-\epsilon_p+
i\Gamma_p/2)(\omega+\epsilon_p+i\Gamma_p/2)},
\end{displaymath}
from which an expression for $G_p(\omega)$ can be obtained. The anomalous
Green's
function $\tilde{G}_p(\omega)$, which is defined by Eq.~(50), is related to
the Green's function $\hat{G}_p(\omega)$ (see definition~(8), and Eqs.~(15)
and (21)) by the formula $\hat{G}_p(\omega)=-2\pi
in_0(T)\delta_p\delta(\omega)+\hat{G}_p(\omega)$.

By comparing between the normal and anomalous Green's functions at
$\omega=\omega_0^s$, we obtain $\tilde{G}_k(\omega_0^s)/G_k(\omega_0^s)\ll 1$
at $\omega_0^s\gg\xi_k$. In this case the
element~(51) of the scattering matrix is determined by the expression
\begin{eqnarray}
T_{k'k}({\bf p}) &=& D_{k}D_{k'}\left[W^*_{q-p,k}W^*_{p-q,k'}
G_{k }(-\omega^s_0)G_{k' }(-\omega^s_0)\widetilde G_{p-q}(\omega-\Omega_-)
\right.
\nonumber\\
&+&\left.
W^*_{-pk'}W^*_{pk}G_{k'}(-\omega^s_0)G_{k}(-\omega^s_0)\widetilde G_{-p}
(\omega-\Omega_-)\right],
\label{53}
\end{eqnarray}
where $G_{k'}(\omega_0^s)\simeq G_k(\omega_0^s)\simeq 1/\omega_0^s$.

The comparison between the latter expression and Eq.~(12) in Section~2
yields $L_{pq}^>$ for the effective Hamiltonian  of exciton recombination:
\begin{equation}
\label{54}
L^>_{pq}=-i\frac{D_qW^*_{pq}}{\omega^s_0}.
\end{equation}
The expressions for the other matrix elements of the exciton
recombination in effective Hamiltonian~(3) can be derived in a similar way.

Thus, the two-photon emission accompanied by the coherent two-exciton
recombination can be analyzed with the aid of the effective Hamiltonian~(3)
with $L_{pq}^>$ determined by Eq.~(54) if the incident light satisfies the
conditions
$|\omega-\Omega_-|\ll\omega_0^s$ and $\xi_k\ll\omega_0^s$. In this case,
the matrix element~(54) does not depend on temperature and ${\bf
f}(\omega_L)$ is identical to the effective matrix element ${\bf F}$
responsible for the phonon-assisted recombination of an isolated
exciton.\cite{20}

\rightline{\bf Appendix B}

\section*{Luminescence intensity of Bose-condensed excitons}

The luminescence of Bose-condensed excitons at frequency $\omega'<\Omega_-$
is due to the optical phonon-assisted excitonic recombination accompanied
by the generation of a
Bogoliubov quasiparticle with energy $\epsilon(p'_L)=\Omega_--\omega'$ in
the excitonic system. The matrix element of this recombination is
\begin{displaymath}
L=L^>_{p'_Lk}v_{p_L}\sqrt{n_{p'_L}+1},
\end{displaymath}

\vspace{-8mm}

\begin{equation}
\label{55}
\end{equation}

\vspace{-8mm}

\begin{displaymath}
v^2_{p'_L}=\frac{\sqrt{\alpha'^2_L+1}-\alpha'_L}{2\alpha'_L},\quad
n_{p'_L}=\left[\exp \left(\frac{|\Delta\omega'|}{T}\right)-1
\right]^{-1},
\end{displaymath}
where $\Delta\omega'=\omega'-\Omega_-$ and
$\alpha'_L=|\Delta\omega'|/\zeta(T)$.

Using Fermi's `golden rule', we obtain the optical phonon-assisted
luminescence intensity $I_s^L(\omega')$ of Bose-condensed excitons at
frequencies $\omega'<\Omega_-$:
\begin{equation}
\label{56}
I^L_s(\omega')=\frac{\omega'^4\sqrt{2m^{3}\zeta(T)
\left(\sqrt{\alpha'^2_L+1}-1\right)}{\bf f}^2(\omega'_L)}
{6\pi^2c^3\sqrt{\alpha'^2_L+1}
\left(\sqrt{\alpha'^2_L+1}+\alpha'_L\right)}
\left[1+{\rm coth}\frac{|\Delta\omega'|}{2T}\right].
\end{equation}

The luminescence of Bose-condensed excitons at frequency $\omega'>\Omega_-$
is due to the optical phonon-assisted excitonic recombination accompanied
by the disappearance of a quasiparticle with energy
$\epsilon(p'_L)=\omega'-\Omega_-$. The matrix element of this recombination
is derived from Eq.~(55) by substituting $v_{p'_L}\sqrt{n_{p'_L}+1}\rightarrow
\sqrt{1+v^2_{p'_L}}\sqrt{n_{p'_L}}$. The luminescence intensity at
frequency $\omega'>\Omega_-$ is derived from Eq.~(56) by substituting
$\alpha'_L\rightarrow -\alpha'_L$ and ${\rm
coth}(|\Delta\omega'|/2T)+1\rightarrow{\rm coth}(|\Delta\omega'|/2T)-1$.
Thus, the expression that determines the luminescence spectrum of
Bose-condensed excitons at an arbitrary frequency $\omega'\ne \Omega_-$ has the
form
\begin{equation}
\label{57}
I^L_s(\omega')=\frac{\omega'^4\sqrt{2m^{3}\zeta(T)
\left(\sqrt{\alpha'^2_L+1}-1\right)}{\bf f}^2(\omega'_L)}
{6\pi^2c^3 \sqrt{\alpha'^2_L+1}\left[\sqrt{\alpha'^2_L+1}-{\rm sign}
(\Delta\omega')\alpha'_L\right]}\left[{\rm sign}(-\Delta\omega')+{\rm
coth}\frac{|\Delta\omega'|}{2T}\right].
\end{equation}

\rightline{\scriptsize Translation was provided by the Russian Editorial office.}

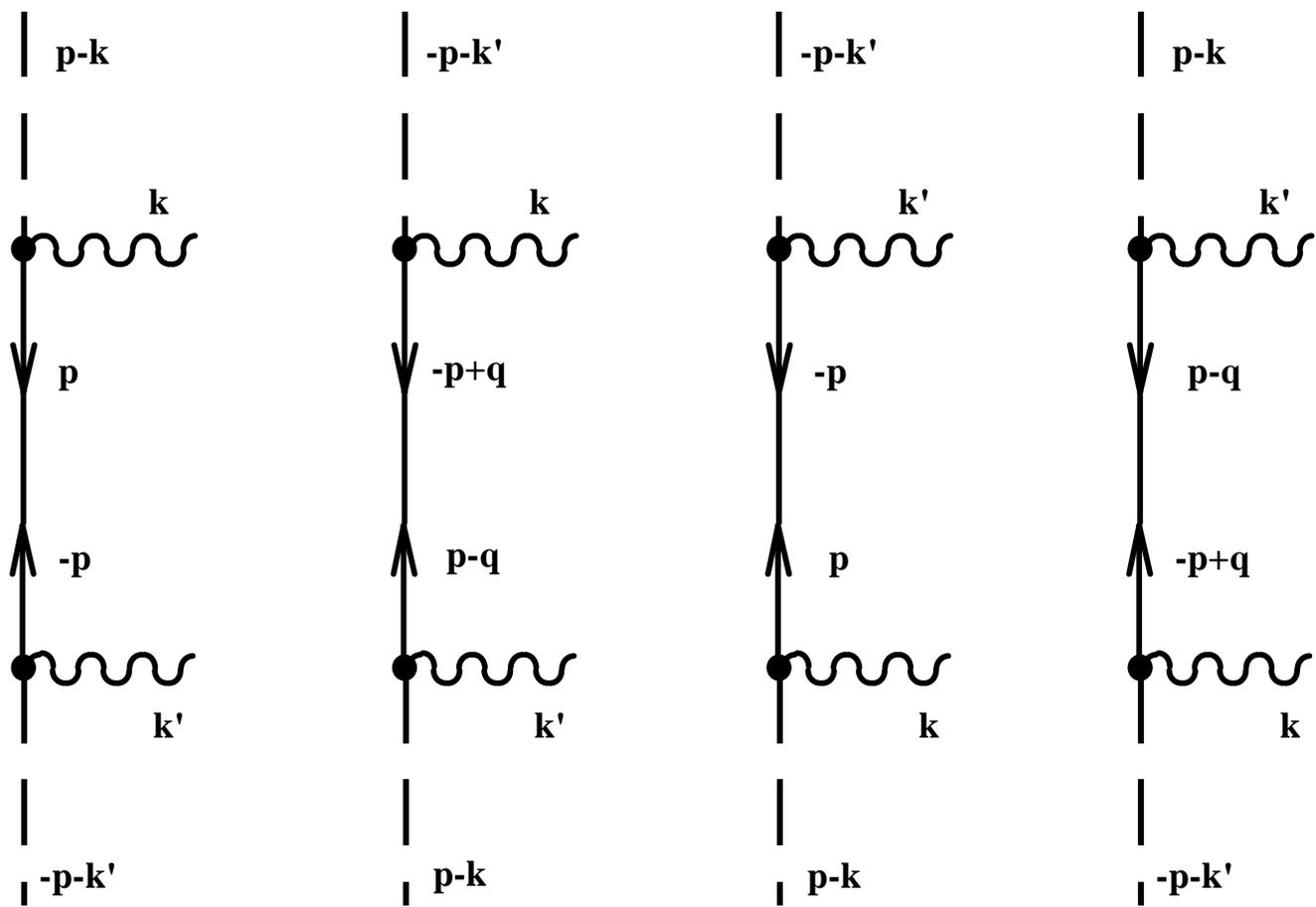
\begin{figure}
\caption{}{Diagrams corresponding to the two-photon emission accompanied by
the coherent phonon-assisted two-exciton recombination (the notations are
explained in the text).\label{Fig1}}
\end{figure}

\begin{figure}
\caption{}{Cross section of the stimulated two-photon emission accompanied
by the coherent optical phonon-assisted
two-exciton recombination:
(a)~as a function of the difference $\Delta\omega=\omega-\Omega_-$ between
the incident light frequency $\omega$ and $\Omega_-$ at different
temperatures of the excitonic system: (1)~$T/T_{\rm c}=0.01$; (2)~0.10;
(3)~0.60; (4)~0.90; (5)~0.99; (b)~as a function of the excitonic system
temperature at different $\Delta\omega$: (1)~$|\Delta\omega|/2T_{\rm
c}=0.2$; (2)~0.3; (3)~0.9. The curves were plotted using Eq.~(30). For all
curves $\mu(0)/2T_{\rm c}=0.3$, and $\tau^L$ is assumed to be
constant.\label{fig2}}
\end{figure}

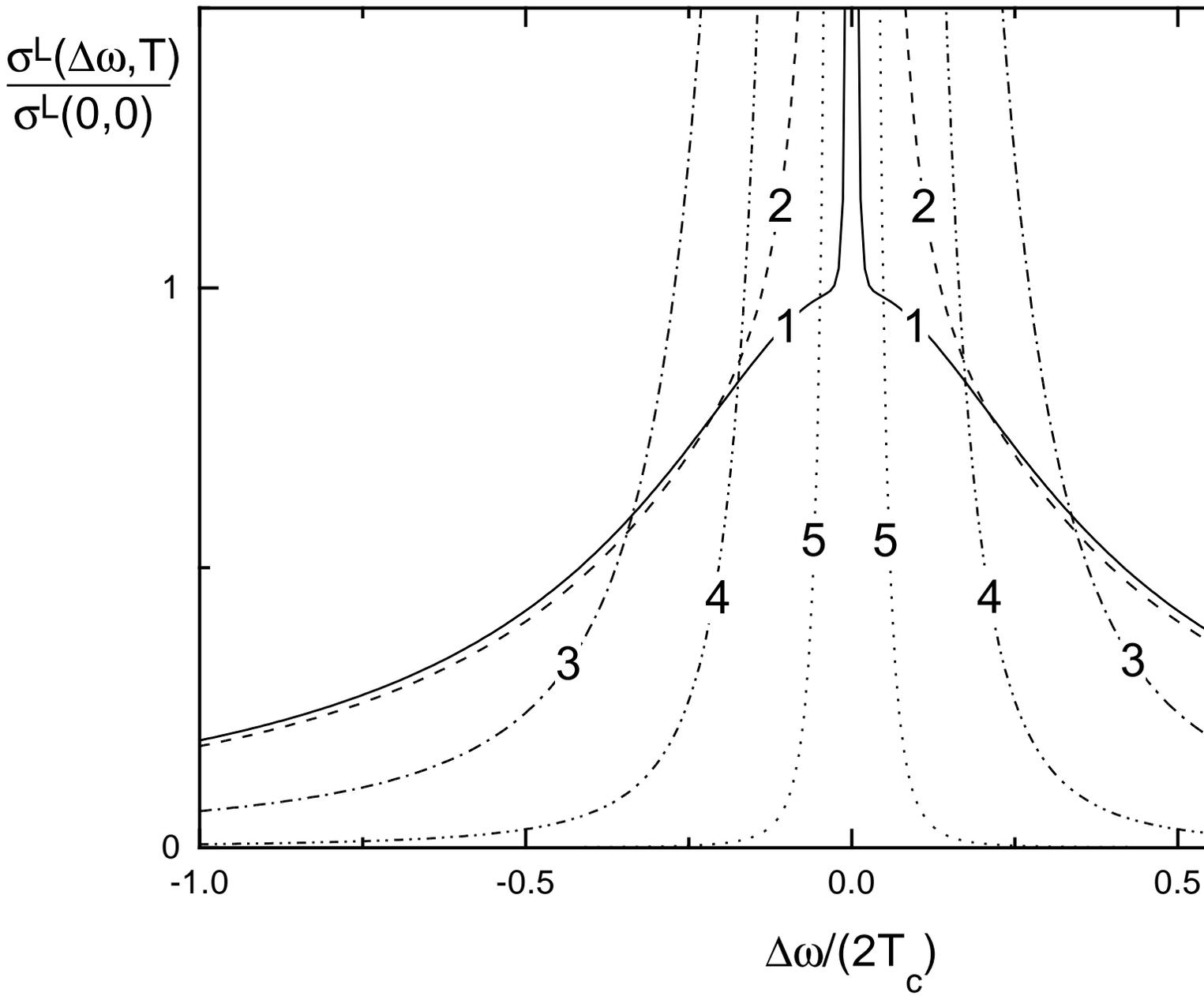
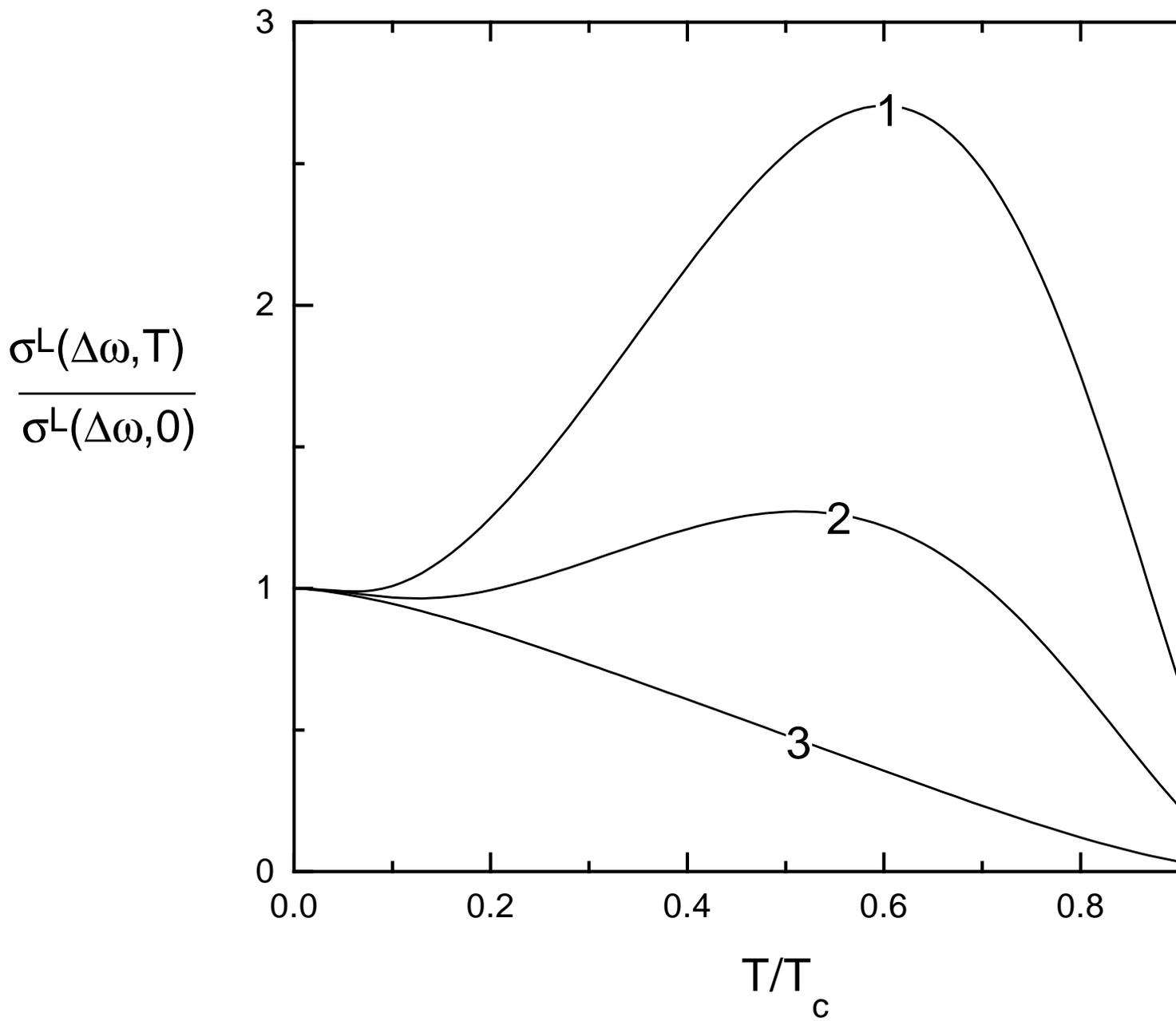
\end{document}